\documentclass[twocolumn]{aastex701}

\usepackage{newtxtext,newtxmath}

\usepackage[T1]{fontenc}

\DeclareRobustCommand{\VAN}[3]{#2}
\let\VANthebibliography\thebibliography
\def\thebibliography{\DeclareRobustCommand{\VAN}[3]{##3}\VANthebibliography}

\usepackage{graphicx}	
\usepackage{amsmath}	
\usepackage{amssymb}	

\usepackage{soul}

\begin{document}

\title[CG splashback]{A Splashback-like Feature of Central Galaxies in Galaxy Clusters}

\author[orcid=0000-0001-5969-4631,sname='Zhang']{Yuanyuan Zhang}
\affiliation{NSF NOIRLab, 950 N. Cherry Ave., Tucson, AZ 85719, USA}
\email[show]{yuanyuanzhang.astro@gmail.com}  

\author[orcid=0000-0002-0298-4432,sname='Bb']{Susmita Adhikari}
\affiliation{Indian Institute of Science Education and Research, Pune, Dr. Homi Bhabha Road, Pashan, 411008, India}
\email{susmita@iiserpune.ac.in} 

\author[orcid=0000-0002-9135-997X,sname='Dd']{Louise O. V. Edwards}
\affiliation{Department of Physics, California Polytechic State University, San Luis Obispo, San Luis Obispo, CA 93405, USA}
\email{ledwar04@calpoly.edu}

\author[orcid=0000-0002-6394-045X,sname='Golden-Marx']{Jesse B. Golden-Marx}
\affiliation{School of Physics and Astronomy, University of Nottingham, Nottingham, NG7 2RD, UK}
\email{jesse.golden-marx@nottingham.ac.uk} 

\author[orcid=0000-0003-2120-1154,sname='Ogando']{Ricardo~L.~C.~Ogando}
\affiliation{Centro de Tecnologia da Informa\c{c}\~ao Renato Archer, Campinas, SP, Brazil - 13069-901}
\affiliation{Observat\'orio Nacional, Rio de Janeiro, RJ, Brazil - 20921-400}
\email{ricardogando@gmail.com} 

\author[0000-0001-9376-3135,sname='Rykoff']{Eli~S.~Rykoff}
\affiliation{SLAC National Accelerator Laboratory, Menlo Park, CA 94025, USA}
\affiliation{Kavli Institute for Particle Astrophysics \& Cosmology,
P.O. Box 2450, Stanford University, Stanford, CA 94305, USA}
\email{erykoff@stanford.edu}




\begin{abstract}
We investigate a splashback-like feature in the outer region of central galaxies (CGs) in clusters. This feature is detected as a "dip" in the radial slope of the CG surface brightness, derived through the stacking of Dark Energy Survey data of over 4000 galaxy clusters in the redshift range of 0.2 to 0.5 with richness 20 and above. The local minimum of the dip occurs between 40 to 60 kpc from the CG center, with a mild dependence on cluster richness. This feature resembles the density transition caused by the splashback effect at the outskirts of galaxy clusters, when accreted matter reaches the apocenter for the first time. We turn to the IllustrisTNG hydro-dynamic simulation to gain theoretical insights. Density bumps, shells and accretion streaks are identified in the diffuse stellar content of the CGs and intra-cluster light which relate to the recent history of disruption and accretion. These features occur at the outskirts of the CGs, up to several hundred kiloparsecs from the cluster center. Thus, the location of the splashback-like dip in the data potentially marks the edge of the CG and the beginning of a region with the cluster diffuse light undergoing active or recent accretion. 
\end{abstract}



\section{Introduction}




Galaxy clusters, the most massive gravitationally-bound structures in the Universe, grow through accretion of matter or galaxy structures. As infalling matter or structures reach the apocenter for the first time, get captured by the galaxy cluster's gravitational potential, and turn-around for the first time, a splashback feature forms \citep{2014JCAP...11..019A, 2014ApJ...789....1D, 2015ApJ...810...36M, Shi:2016lwp}. 
Studies of simulations and observations have found that the average splashback radius can be close to the galaxy cluster's $R_{200m}$, which is the radius within which the galaxy cluster's average density is 200 times the Universe mean. The splashback radius can depend on the cluster's accretion rate \citep{2014ApJ...789....1D},  and often corresponds to a local minimum in the slope, or radial gradient, of the cluster's density profiles \citep{2017ApJ...841...34M, 2017ApJ...843..140D}. Thus, the splashback radius can be measured 
by using weak gravitational lensing mass density measurements \citep{2017ApJ...836..231U,2018ApJ...864...83C, 2019MNRAS.485..408C, 2019MNRAS.487.2900S} and galaxy density measurements \citep{2016ApJ...825...39M, 2017ApJ...841...18B}. These measurements provide additional information that helps us understand the cluster's accretion process and how galaxies evolve in the cluster environment \citep{ 2021ApJ...923...37A, 2021ApJ...911..136B}, as well as a new way to estimate cluster mass \citep{2025ApJ...988..149G}. 


Recently, using simulations, \cite{2021MNRAS.500.4181D} proposed that the splashback radius can be measured through investigating the radial density distribution of cluster's diffuse stellar light, also known as intra-cluster light (ICL). This potential has been further confirmed in simulation studies by \cite{2025arXiv250802837D} and \cite{2025arXiv250807232W}. Similar to the cluster's weak lensing mass measurements or galaxy density measurements, one may detect a local minimum when examining the slope of the diffuse light's density, corresponding to the cluster's splashback radius. Observing splashback through stellar light can help us further constrain the galaxy cluster's accretion process, and understand the evolution of the cluster's baryonic components. 


However, in an attempt to measure this splashback feature by examining one galaxy cluster, \cite{2021MNRAS.507..963G} discovered not just one, but two local minima in the slope of its ICL radial profile. While the outer minimum lies towards the cluster outskirts, which may correspond to the cluster's splashback feature (although at a radius smaller than likely predicted by splashback models), the other minimum appears at a radius of $\sim $ 70 kpc, which is at the outskirts of the cluster's central galaxy (CG). \citet{2021MNRAS.507..963G}  speculated that the inner minimum  may correspond to the splashback feature of the CG itself, and indicate a transition from the CG to the ICL. 

In this paper, we follow the path of \cite{2021MNRAS.500.4181D} and \cite{2021MNRAS.507..963G} to search for the splashback signature using ICL measurements of over 4000 galaxy clusters, based on one of the largest galaxy cluster samples used for ICL studies \citep{2024MNRAS.531..510Z}. Our goal is to acquire high signal-to-noise measurements of local minima in the slope profile of CG and ICL, and we confirm a splashback-like feature at the outskirts of the CG, in the CG to ICL transition region.

The rest of this paper is organized as follows. We briefly review the measurements used in this analysis in Section~\ref{sec:data}.  Section~\ref{sec:results} contains our main result of a gradient transition feature in the outer region of the CG as well as the robustness analysis of the result. Section~\ref{sec:sims} describes a simulation analysis in which we investigate the accretion features at the CG outskirts in the hydrodynamic IllustrisTNG simulation. Finally, Section~\ref{sec:discussion} discusses and summarizes our analyses. Throughout this paper, we assume a flat $\Lambda$CDM cosmology model with $\Omega_m = 0.30$ and $h=0.7$.

\section{Data and Method}\label{sec:data}

This analysis is based on the ICL surface brightness measurements presented in \cite{2024MNRAS.531..510Z}, which use the full six years of imaging data from the Dark Energy Survey (DES), with a sample of over 4000 redMaPPer galaxy clusters \citep{2014ApJ...785..104R} in the redshift range of 0.2 to 0.5. \cite{2024MNRAS.531..510Z} presents the averaged measurement of CG and ICL radial profiles for galaxy clusters in richness and redshift bins, which is the basis for the slope measurements presented in the next section. Here, we briefly summarize their data and measurements:
\begin{itemize}
\item Wide-field optical images in $g$, $r$, $i$, $z$ and $Y$-bands were taken by the Dark Energy Camera \citep{2015AJ....150..150F} for DES, from 2012 to 2019 \citep{DES:2019rtl}. Single exposure images were processed by the DES data management team. Critically, the sky background for each raw image is estimated  over the full field-of-view (FOV) using a principle component analysis (PCA) method \citep{2017PASP..129k4502B}, and subtracted from the processed images. The measurements were based on processed single epoch  $r$, $i$ and $z$-band images. 
\item The DES collaboration has publicly released processed single epoch images, coadd images, coadd catalogs as well as quality assurance maps \citep{2018ApJS..239...18A, 2021ApJS..255...20A}, and quality-improved "gold" catalogs \citep{2021ApJS..254...24S}. Nominally, the coadded DES Year 6 catalog reaches 23.9, 23.4 and 22.7 magnitude respectively in $r$, $i$ and $z-$bands at $S/N=10$ for extended objects \citep{2025arXiv250105739B}.
\item The redMaPPer galaxy cluster finder has been applied to the DES Year 3 gold coadd catalogs (based on coadding DES Year 1 to Year 3 observations)\footnote{\url{https://des.ncsa.illinois.edu/releases/y3a2/Y3key-cluster}}, yielding a high-quality galaxy cluster catalog that is also used for cosmological analysis \citep{2025arXiv250313632D}. \cite{2024MNRAS.531..510Z} used version 6.4.22+2 of the DES Year 3 redMaPPer cluster catalog, with a richness threshold of 20 and redshift range of 0.2 to 0.5, to study ICL. The CG is determined by the redMaPPer algorithm, and multi-wavelength studies \citep{2015MNRAS.454.2305S, 2019MNRAS.487.2578Z, 2020ApJS..247...25B, 2024MNRAS.533..572K} estimated that the redMaPPer algorithm has an $\sim 80\%$ success rate of finding the galaxy that corresponds to the central galaxy of the cluster dark matter halo  \citep[as often defined in simulation studies of cluster-sized dark matter halos][]{2007MNRAS.375....2D, 2008ApJ...676..248Y}.
\item For each galaxy cluster, the DES single-epoch images containing the galaxy clusters are coadded together using the Swarp software \citep{2002ASPC..281..228B}. Note that additional sky background subtraction may be enabled during the Swarp coaddition process, but for deriving the products in \cite{2024MNRAS.531..510Z}, the background subtraction option was disabled. 
\item Galaxies or point source objects identified in the DES Y6 coadd catalogs (derived from coadding six years of DES observations) above a cluster-redshift-dependent magnitude limit, excluding the CG, were masked out of the images. 
\item From the masked image, a surface brightness profile of the CG and ICL was derived in circular radial bins, and then averaged over a sample of clusters to derive raw stacked surface brightness measurements. 
\item Finally, the same stacking procedure was applied to a set of random sky locations that track the sky coverage of the redMaPPer catalog. This average surface brightness profile was subtracted from the cluster's stacked measurements to derive a final, combined measurement of the cluster's CG and ICL profile (referred to as CG+ICL or diffuse light in this paper). 
\item Uncertainty estimations in \cite{2024MNRAS.531..510Z} were derived with a Jackknife sampling method \citep[e.g.,][]{2009MNRAS.396...19N, 2017MNRAS.469.4899M}, by dividing the DES Y6 sky coverage into 50 patches, and sampling the measurements from those 50 patches. This sampling method is often used in cosmological analysis to capture both the intrinsic measurement noise, as well as noises caused by large-scale structures.
\end{itemize}

The final products we use from \cite{2024MNRAS.531..510Z} are the surface brightness profiles of CG+ICL of galaxy clusters in different richness and redshift bins. We also estimate measurement uncertainties using the Jackknife sampling method described above to capture both measurement noises and large-scale variations across the whole DES footprint.

\section{Slope and Dip Measurements}\label{sec:results}

\begin{figure}
	\includegraphics[width=1.0\columnwidth]{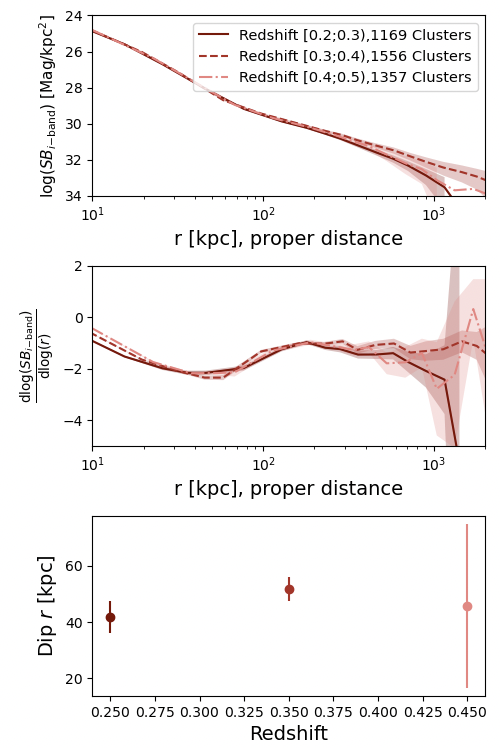}
    \caption{Upper panel: Surface brightness profiles of redMaPPer clusters in three redshift bins. These profiles were presented in \cite{2024MNRAS.531..510Z}, each consisting of over 1000 stacked clusters. Middle: Derived slopes of the surface brightness profiles in different redshift bins. Note the dip feature between 20 and 100 kpc in all of the profiles at the CG outskirts. Bottom: the local minima of the dips are identified to be between 42 to 52 kpc, and do not seem to have significant dependence on cluster redshift.}
    \label{fig:redshift}
\end{figure}

\begin{figure*}
    \includegraphics[width=2.0\columnwidth]{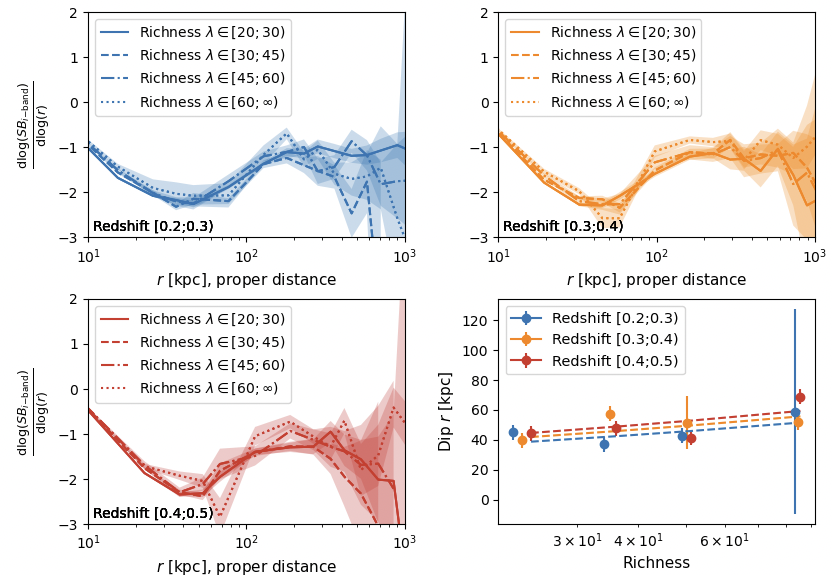}
    \caption{Upper left, right and lower left panels: Slope profiles of redMaPPer clusters in different redshift and richness bins. The dip features are present in all cluster subsets. Lower right panel: the local minima of the dips in different redshift and richness subsets. We detect a richness-dependence in the minimum radii, but no significant redshift evolution. The dashed lines show the fitted linear relations between the minimum radii and cluster richness at their median redshifts.}
    \label{fig:richness_redshift}
\end{figure*}

Using the radial profile of the CG+ICL surface brightness from \cite{2024MNRAS.531..510Z}, we further derive their radial gradients at the logarithmic scale, which are similarly defined with the radial gradient measurements used in galaxy cluster splashback measurements \citep[e.g., ][]{2018ApJ...864...83C, 2021ApJ...923...37A}. This method can be particularly efficient at revealing features of transitions and density shocks. Specifically, the radial gradient is defined as:
\begin{equation} 
g(r) = \mathrm{d log_{10}} S(r) / \mathrm{d log_{10} (r)}.
\end{equation}
In the above equation, $S(r)$ is the CG+ICL surface brightness radial profile and $r$ is the corresponding radius. 
For the rest of this paper, we will also refer to the radial gradient profile as the ``slope'' measurement. 

\subsection{A Dip at the CG outskirts}

Figure~\ref{fig:redshift} shows the surface brightness and gradient profiles of the diffuse light in three cluster redshift samples. In these profiles, we first observe that the CG radial profiles steepen at a larger radius (increasingly negative slope) within 30 kpc. Then, between the 30 to 100 kpc radial range, the slope reaches a local minimum, and becomes less steep with increasing radius. Outside of 200 kpc, the slope appears to flatten.

 We find it particularly intriguing that the radial gradients have a local minimum, or "dip", between 30 to 100 kpc. A similar effect has been noted in \cite{2021MNRAS.507..963G} (see Section~\ref{sec:discussion} for a detailed discussion). The dip resembles the splashback feature corresponding to the first apocenter reached by infalling matter or substructure, as described in the introduction. If this dip has a similar nature to the cluster splashback phenomenon, it could indicate a transition or accretion region, located in the outer region of the cluster CGs. 

 To quantitatively locate the dip, we first smooth the surface brightness profiles $S(r)$, with a Savitzky-Golay filter \citep[e.g., as in][]{2016ApJ...825...39M} (filter window length 5 and polynomial order 3) and derive gradient profiles. \footnote{The smoothing filters out noisy peaks in the profiles and makes it easier to identify transition features in the gradient profiles. The smoothing is not necessary for the stacked gradient measurements, but helps reducing noisy measurements in the simulation analyses in later sections. For consistency, we apply the smoothing to both data and simulation-based measurements.} We then find the minimum in each smoothed gradient profile between 10 and 100 kpc, which is between 42 to 52 kpc for the three redshift bins shown in Figure~\ref{fig:redshift}. As previously mentioned, the uncertainties of the dip location are estimated using a Jackknife sampling method and are generally estimated to be between 1 to 10 kpc. However, we find that the uncertainties can be severely under-estimated (sometimes with zero variance between Jackknife samples) when we use large radial bins. To avoid dependence on the radial bin size, we assume a minimum measurement uncertainty of 5 kpc to avoid the under-estimation.

Interestingly, the dip locations are qualitatively comparable to radius of the CGs (depending on the  definition of "radius", measurements vary). For example, \cite{2020ApJS..247...43K} have measured the median effective radius of CGs to be around  45 or 74 kpc for BCGs with light profiles well fitted by a single Sersic function or double Sersic functions respectively. Given the intriguing location of the dip features, we speculate that they may be related to the mechanisms that physically separate CG and ICL.


\subsection{Relationships between Cluster Properties and the Dips}

We further investigate how the location of the dip changes with the mass and redshift of the cluster, by sub-dividing the cluster samples into richness and redshift subsets and remeasuring the location of the dips. As we do not have direct cluster mass estimates, we make use of redMaPPer's richness quantity, which is a probabilistic count of red member galaxies inside the redMaPPer-identified clusters. Previous studies have quantified the scaling relation between richness and cluster masses measured through multi-wavelength observations and weak lensing \citep[e.g.,][]{2019MNRAS.482.1352M, 2019MNRAS.490.3341F}. 
Further information about these richness and redshift subsets can be found in \citep{2024MNRAS.531..510Z}. Here, we analyze the dependence of dip measurements on cluster richness and redshifts, and interpret richness dependence (if any) as cluster mass dependence. 


Figure~\ref{fig:richness_redshift} shows the slope profile and dip location for the cluster richness and redshift subsamples. The measurements are noisier than those presented in Figure~\ref{fig:redshift}. Nevertheless, dips are present in all of the profiles. We locate the minima of these dips using the same method described above, which varies between 40 to 70 kpc.

We do not notice any redshift evolution in the dip minimum, but there appears to be a subtle trend of the minimum increasing with cluster richness. To quantify the dependence, we fit the dip radius measurements to the equation below (using the EMCEE package and a Gaussian likelihood assumption that accounts for measurement uncertainties). 
\begin{equation}
    R=R_0\times (\frac{1+z_\mathrm{bin}}{1+0.25})^a \times (\frac{\lambda_\mathrm{bin}}{25.0})^b 
\label{eq:richness}
\end{equation}
We find that the fitted model favors a richness dependence with a $b$ value of $0.229\pm0.073$ (above 0 at a $3.1\sigma$ level) but no significant redshift dependence ($a=0.94\pm0.59$).  However, we note that the detection for a richness-dependent dip location may need to be verified by more precise measurements. The detection of a positive $b$ value here is especially sensitive to the measurement uncertainty of the minimum radius: if assuming minimum measurement uncertainties of 10 kpc (instead of 5 kpc), the fitted $b$ value becomes $0.22\pm0.15$, only deviating from 0 at a $1.5\sigma$ level. Nevertheless, a richness-dependent dip radius is not surprising: in Figure ~\ref{fig:richness_redshift}, the radial slope of the CG between 10 kpc and 30 kpc appears to be slightly less steep in richer clusters, which pushes the dip radius slightly outward.

\subsection{Robustness of the Dip Detection}

\begin{figure}
	\includegraphics[width=1.0\columnwidth]{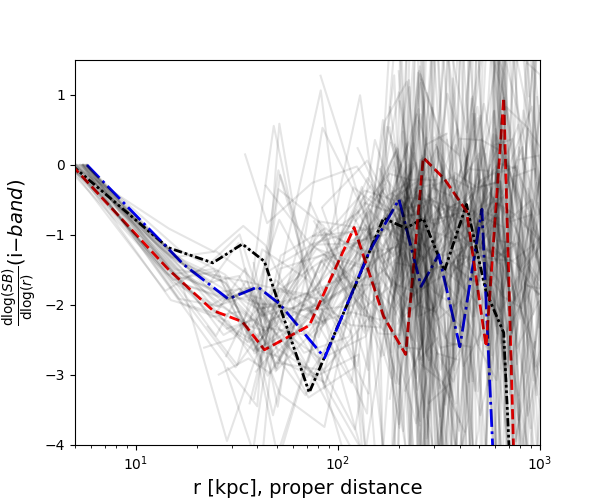}
    \caption{Slope profiles of individual redMaPPer clusters. The profiles of three clusters are highlighted in this figure to demonstrate the trends of individual clusters. While the CG slope rapidly steepens (more negative) at larger radius within the first 30 kpc, the profiles become very noisy afterwards. It is much more difficult to identify the dip  in the individual profile measurements. }
    \label{fig:individuals}
\end{figure}

\begin{figure}
    \includegraphics[width=1.0\columnwidth]{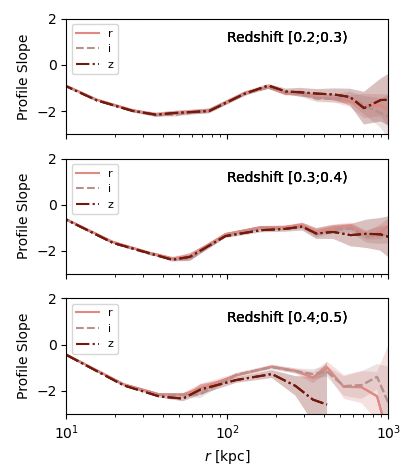}
    \caption{Slope profiles of redMaPPer clusters with surface brightness measurements in different optical bands, $r$, $i$ and $z$. The slope measurements within 200 kpc and the dip locations are highly consistent among measurements in different bands. }
    \label{fig:bands}
\end{figure}

The results presented in the previous subsections are based on stacked diffuse light measurements. Is stacking necessary for detecting the dips in this analysis? Here, we examine the slope profile of individual galaxy clusters (these individual clusters are also studied in \citealt{2025MNRAS.538..622G}), rather than the slope profiles of stacked galaxy clusters as presented in previous subsections. 

We derive the slope profiles for the galaxy clusters in the redshift 0.3 to 0.4 slice, and present them in Figure~\ref{fig:individuals}. Generally, these slope profiles do appear to drop within 100 kpc, and then rise again to reach a plateau at a larger radius (several hundreds of kpcs). However, the individual measurements are much noisier. There can be several dips in the individual profiles outside of the immediate CG center ($\sim$20 kpc), making it difficult to distinguish between a real dip feature and noise in the data. Additionally, as previously discussed in \cite{2017ApJ...843..140D, 2017ApJ...841...34M}, detecting a splashback feature in a single system based on density profiles is difficult, and these measurements can be affected by measurement noise, dense environments, massive substructures or large neighboring groups and clusters. Nevertheless, we are encouraged that the trends in the individual profiles are consistent with the trends in the stacked profile.

We examine a few additional aspects of the stacked measurements to verify the robustness of the results. 
The default results presented in this paper are based on DES $i$ band observations, which cover the post-4000\,\r{A} break for the entire redshift range. It is also deeper than the $z$ band available from DES. To test the robustness of the dip feature, we perform the same slope measurements in DECam $r$ and $z$  bands as shown in Figure~\ref{fig:bands}. The gradient profiles and dip feature are highly consistent across different bands.  

Another factor we have considered is how the Point Spread Function (PSF) may have influenced the slope  or the dip measurements. We experimented with convolving a stacked simulation profile (from TNG300-1, described in the next section) with PSF models, and find that the PSF only affects the gradient measurements within $\sim$20 kpc up to reshift 0.5. The effect varies with redshift because of the changing angular size of CGs on the sky. However, the PSF itself does not create a dip feature as the feature is not present in the pre-PSF-convoluted simulation profile. In Figure~\ref{fig:redshift}, we can already observe the PSF effect where the measured CG slope becomes steeper with increasing redshift within 30 kpc, but there is no consistent redshift-dependent trend in the gradient profiles around the dip minima.

\section{Simulation Interpretation} \label{sec:sims}

To gain insight into the physical origin of the dip, we analyze the evolutionary history of stellar material in the IllustrisTNG simulations\footnote{https://www.tng-project.org} \citep{2019ComAC...6....2N, 2018MNRAS.475..648P, 2018MNRAS.475..676S, 2018MNRAS.475..624N, 2018MNRAS.477.1206N, 2018MNRAS.480.5113M}, which have previously been used to study diffuse stellar and ICL \citep[e.g.,][]{2021MNRAS.501.1300S, 2023MNRAS.521..478G, 2024MNRAS.529.4666A, 2025arXiv250616645M, 2025ApJ...988..229Y, 2025MNRAS.tmp.1059M}. Below, we briefly overview the simulation products used. 
\begin{itemize}
\item We use the TNG50-1 hydrodynamic simulation \citep{2019MNRAS.490.3234N, 2019MNRAS.490.3196P} to analyze the evolution of the stellar content of one galaxy cluster-sized halo. TNG50-1 has the highest resolution in the IllustrisTNG simulation suite. We use this simulation expecting more realistic stellar properties which is important to the study here. We also performed similar analyses with TNG50-2 and TNG50-3, which have lower resolutions. TNG50-1 has the sharpest stellar accretion/disruption features among the three. 
\item The analyses are primarily based on the stellar particles in the hydro simulations. The minimum baryon particle mass in the TNG50-1 simulation is $5.7\times 10^{4} \mathrm{M_{\odot}}/h$. Other stellar quantities that are relevant to this work are their Coordinates, Velocities and ParticleIDs. For example, the radial velocities of the stellar particles are calculated from their velocities projected along the halo center direction. The ParticleIDs quantity  is used to match and track the particles in different redshift snapshots. 
\item The simulation subhalo properties are provided in the subfind subhalos table of the simulations. We use the subhalo's SubhaloHalfmassRad quantity, which is a radius measurement of a subhalo containing half of its total mass, to determine if a stellar particle is associated with a subhalo. If a stellar particle falls within 3 times this radius of a subhalo, then we consider this stellar particle to be associated with a subhalo. Note that under this criteria, a stellar particle can be associated to several subhalos. We use all subhalos, except the first one of a group (which is the CG itself) to determine subhalo association. Stellar particles not close enough to any subhalos (except the CG itself) are considered to be a particle of the CG+ICL component, which we also refer to as a diffuse stellar particle in this paper.
\item We make use of both the TNG50-1 full snapshots and also the subbox snapshots. We use the full snapshots to analyze whether or not the stellar particles are associated with subhalos at $z>0$. The subbox snapshots are centered on the most massive halo in TNG50-1 and allow us to track the detailed evolutionary history of the stellar particles with fine redshift resolution.
\item Finally, to study the stacked density profiles of galaxy clusters in simulations, we also use the TNG300-1 simulation. This simulation has a baryon particle mass resolution of $7.6\times 10^{6}  \mathrm{M_{\odot}}/h$, much higher than the TNG50-1 simulation, but contains many more cluster-sized halos because of its larger volume, allowing us to acquire more precise measurements of cluster stellar density profiles after stacking.
\end{itemize}

\subsection{The Story of One Cluster}


In this section, we use the TNG50-1 simulation and focus on the stellar evolution history of the most massive dark matter halo in the simulation. This halo has a mass of $10^{14.23} \mathrm{M_\odot}/h$ at redshift $z=0$. 
We examine the spatial locations and velocities of the stellar particles for any signatures of splashback in their history. 

\subsubsection{Profile and Phase Space Diagram at $z=0$}

\begin{figure*}
    \includegraphics[width=2.1\columnwidth]{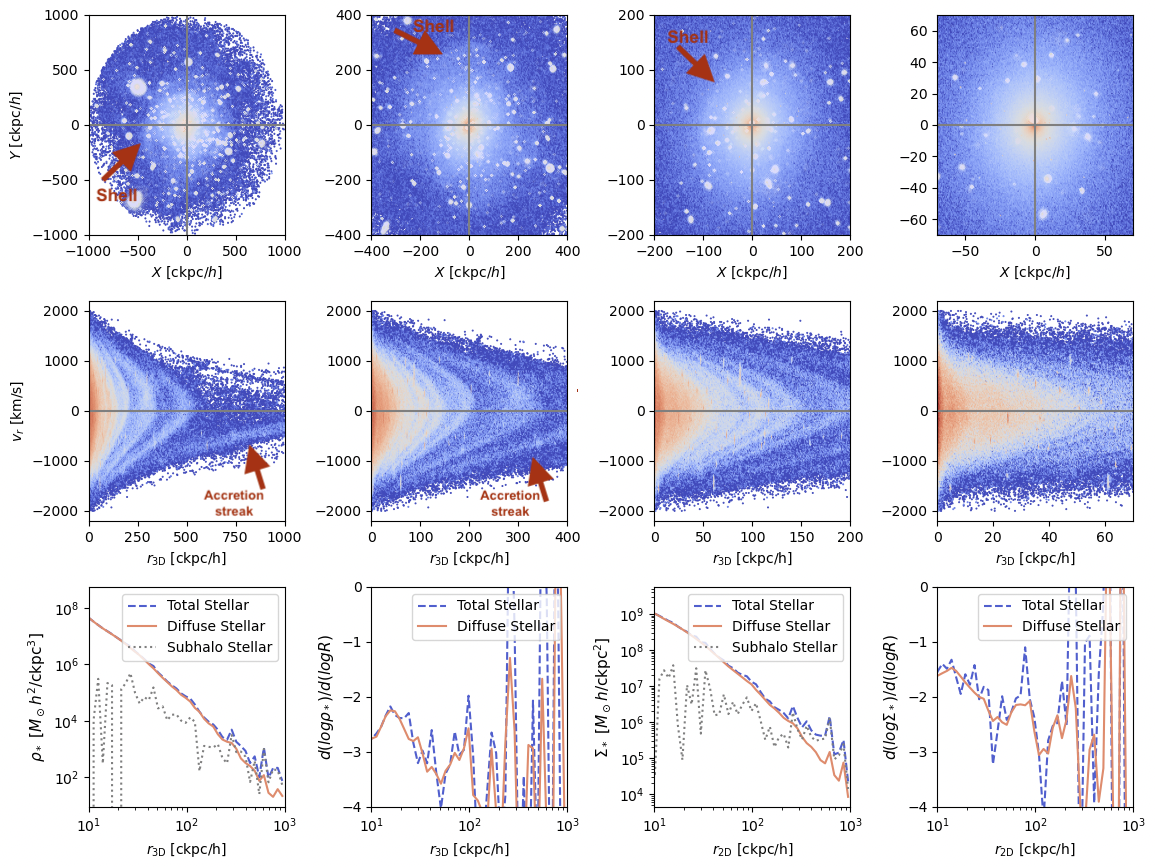}
    \caption{Upper row: the stellar density maps of one cluster-sized halo in the TNG50-1 simulation, projected onto the $X-Y$ plane, zoomed in to show details at different radial scales from left to right. Shell-like density bumps are evident in these density maps, for example, around 500 ckpc$/h$, 300 ckpc$/h$ and 100 ckpc$/h$. The stellar particles associated with subhalos are grouped in white blobs in these figures. Middle row: Phase-space diagram (radial velocity {\it vs} radial distance) of the diffuse stellar particles (excluding the subhalos shown in the upper row), zoomed in to show details at different radial scales from left to right. The shell-like structures become more prominent in this diagram. Lower row: 3D (1st panel) and 2D-projected density profiles (3rd panel) of the cluster stellar contents, and their radial slopes (2nd and 4th panels). Shell-like density bumps in the density and phase-space diagrams often correspond to dips and bumps in the density profiles.}
    \label{fig:phase_space}
\end{figure*}

We first examine the halo's stellar particle properties at $z=0$ in Figure ~\ref{fig:phase_space}. The first row shows the halo's stellar density, projected onto the $X-Y$ plane of the simulation, with progressive zoom-ins from left to right. We separate the CG+ICL component and the cluster's subhalo stellar content (the white blobs in the figure). A very extended diffuse stellar component is obvious in this cluster, going to the radius limit of the particle selection at 1000 ckpc/$h$ (The letter c in the unit ckpc/$h$ refers to comoving distance). On top of the very extended diffuse component, the density maps also show shell-like features that indicate past galaxy disruption or stripping events \citep[also see simulation studies in][]{2018MNRAS.480.1715P, 2024MNRAS.530.4422K}, for example,  around 500 ckpc/$h$ (1st panel of the first row), 300 ckpc/$h$ (2nd panel of the first row) and around 100 ckpc/$h$ (3rd panel of the first row). We also compute the radial stellar density profiles of the cluster, in terms of 3D distance (1st panel in the 3rd row) or 2D projected distance on the $X-Y$ plane (3rd in the 3rd row). These shell features present in the density maps generally correspond to caustic-like features in the radial density profiles (due to the accumulation of star particles at apocenters). When examining the slopes of the density profiles, these features further produce localized density minima in the smoothed logarithmic slope profiles (2nd and 4th panels in the 3rd row), for example around 100 ckpc/$h$ and 300 ckpc/$h$ respectively (The local minima between 10 and 500 ckpc/$h$ in the smoothed $\Sigma_*$ profiles are at 51.19, 121.59, 187.39,
 321.77, and 445.07 ckpc/$h$. 

As performed in \cite{2014JCAP...11..019A} and \cite{2020MNRAS.493.2765S}, we examine the radial velocity vs radial distance phase space of the diffuse stellar particles, excluding those in subhalos (2nd row of Figure~\ref{fig:phase_space}). The phase space diagram is more effective at revealing these shell-like features, with a first major outer shell feature around 500 ckpc/$h$, and a second inner shell feature around 300 ckpc/$h$, corresponding well to the density enhancements discussed above. When zooming into small radial scales, the phase-space diagram reveals many more shell features within 200 ckpc/$h$, which are prevalent at small radial scales, and can be seen as close as 40 ckpc/$h$ to the center of the CG. When examining the history of the particles in these shell structures (next subsection), we notice that the stellar particles in the same shell are often deposited by the same sub-halo, which explains why they occupy similar orbits around the CG.

Notably, we also find an accretion streak/infall stream of the diffuse stellar particles, similar to the simulation study in \cite{2025arXiv250807232W}. \cite{2014JCAP...11..019A} and \cite{2020MNRAS.493.2765S} have shown that a streak in the negative radial velocity space, with no counterpart in the positive velocity half, indicate accreting structures on first infall that have not settled into an an orbital track. In the first panel showing the cluster phase-space diagram between 0 and 1000 ckpc$/h$, we notice a prominent accretion streak that spans the radial range of 100 ckpc$/h$ to 1000 ckpc$/h$. Between 500 and 1000 ckpc$/h$, this negative accretion streak dominates the cluster diffuse stellar component, with 76.7\% of the diffuse stellar particles having negative radial velocities. However, the contribution of this streak to the total diffuse mass is minor compared to the diffuse stellar particles forming shells in the phase-space diagram. Given their location in the phase space diagram, it is implied that these diffuse stellar particles are the leading or trailing arm of tidally dissociating sub-halos during infall into the cluster potential well.

To summarize, by examining the halo's stellar density distribution and phase space diagram at $z=0$, we identify (1) shell structures that indicate recent disruption or accretion events, and (2) an accretion component that is likely already diffuse before their accretion into the cluster potential well. The first component can create small density bumps in the radial stellar density profiles as well as dips in the slope profile. 



\subsubsection{Stellar Particle history}

\begin{figure*}
    \includegraphics[width=2.0\columnwidth]{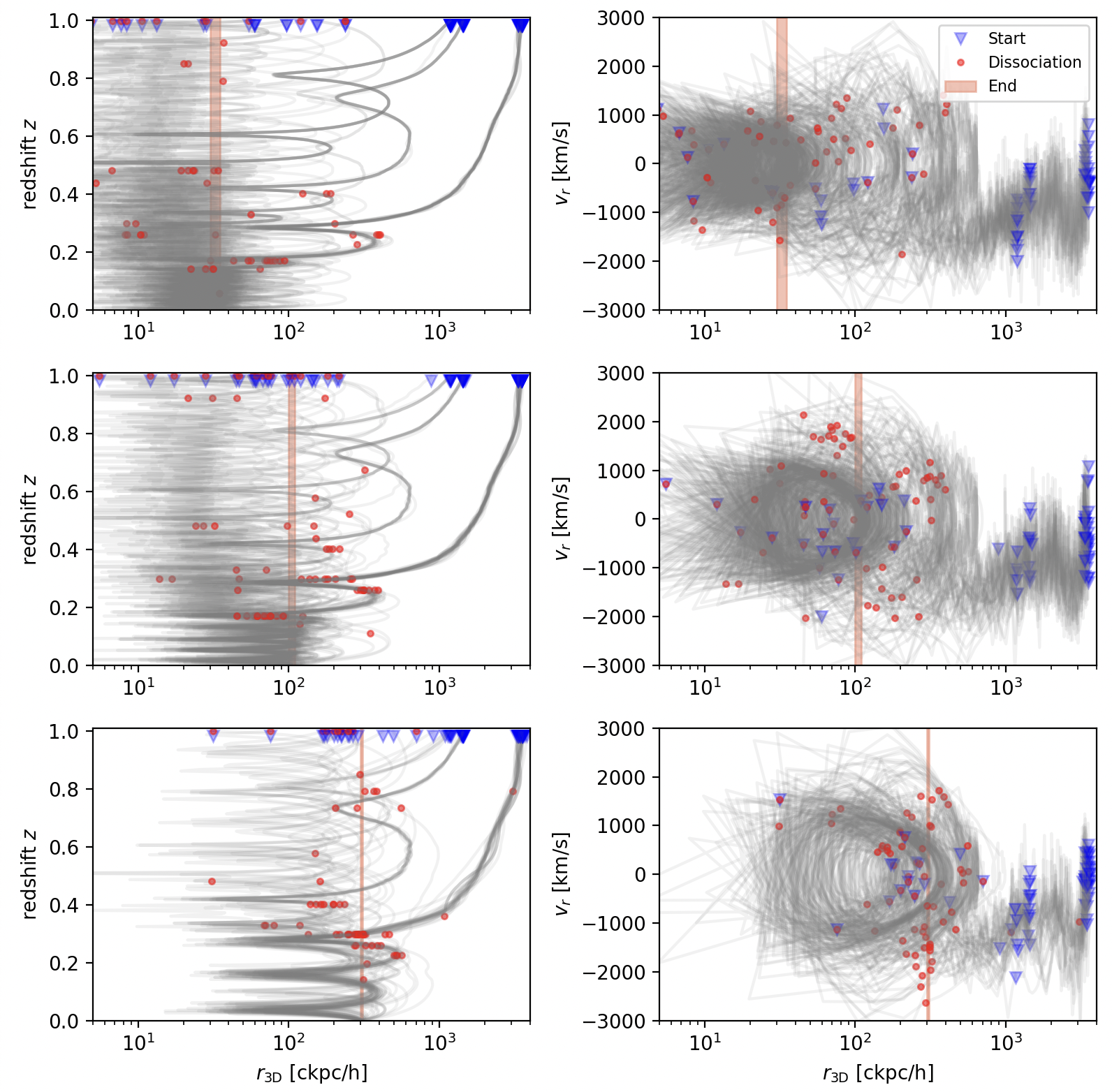}
    \caption{Evolution history of the stellar particles in the TNG50-1 simulation. Upper row: the history of a sample of diffuse stellar particles in the radial range of [30, 35] ckpc$/h$ at $z=0$. The upper left panel shows the evolution of their radial distances to the cluster center. Some of the particles have settled into close orbits as early as $z=1$. Others started as far away as $>3000$ ckpc$/h$. The upper right panel shows the evolution of the particles' phase space diagram. For the particles that traveled great radial distances from $z=1$ to $z=0$, they tend to start on an accretion streak with negative radial velocities, then make a couple of orbits at an intermediate distance with high radial velocities, and finally settle into an orbit close to the cluster center.  The blue triangles and pink bands respectively mark the particles' initial locations at $z=1$, and final locations at $z=0$. The orange circles mark the approximate locations when they become dissociated from a subhalo (outside 3 times SubhaloHalfmassRad of subhalos), or their initial locations if they are already dissociated at redshift $z=1.0$.  Middle and lower rows: same figures but for diffuse stellar particles in the radial range of  [100, 110] ckpc$/h$, and [300, 310] ckpc$/h$ at redshift 0. There are similar evolution trends with the diffuse stellar particles in the radial range of [30, 35] ckpc/$h$. The bottom panels also contain particles with high angular momentum, whose orbits  do not plunge to small radii.}
    \label{fig:stellar_history}
\end{figure*}


In this subsection, we examine the history of a sample of diffuse stellar particles and how their locations and velocities have changed with redshift. We identify a few ($\sim 300$) stellar particles that are no longer associated with subhalos at $z=0$ (those not in the white subhalo structures in Figure~\ref{fig:phase_space}) and track their radial distances and velocities through matching their particle IDs in different redshift snapshots since $z=1$. We also determine when these particles become dissociated from subhalos, through tracking their distances to subhalos in different redshift snapshots, and whether or not these distances have exceeded 3 times the radius of the subhalos\footnote{We use the subhalos' SubhaloHalfmassRad quantity as their radius.}. 

Figure~\ref{fig:stellar_history} shows the history of particles with their final radial distances in three radial ranges [30, 35] ckpc/$h$, [100, 110] ckpc/$h$, and [300, 310] ckpc/$h$ at $z=0$. 
These three radial ranges are selected to be close to the CG, in the CG to ICL transition region, and in the ICL. 

We first focus on the history of the stellar particles with their $z=0$ distance in the [30, 35] ckpc/$h$ radial range. The top left panel of Figure~\ref{fig:stellar_history} shows how the radial distance to the CG center evolves with redshift for the stellar particles. Some of the particles are already close to the CG center and have fallen in earlier than $z=1$. Others are still bound to subhalos at $z=1$ and have only recently ventured into the cluster center. This includes a few particles that have radial distances further than 500 ckpc/$h$ at $z=1$, some as far as $\sim$3000 ckpc/$h$! These initially-distant particles first make a couple of orbits in an intermediate radial range (with a radius of several hundred ckpcs/$h$) during which they often become dissociated from subhalos. After these initial orbits, they eventually settle into orbits much closer to their $z=0$ radius (within 100 ckpc/$h$). The right panels of Figure~\ref{fig:stellar_history} show the trajectories of the same particles in phase space, $v_r-r$.

For particles that fall in along with subhalos starting at a distance of $r>3000$ ckpc/$h$ at $z=1$, we clearly see that they fall in along the accretion streak in phase space, with negative radial velocities as noted in the previous subsection. These star particles usually appear to dissociate from their subhalos, between their closest passage to the center and subsequent apocenter. The dissociation points are marked by red dots in the figure. For radially plunging orbits, dissociation points appear at the first pericenter and splashback (apocenter where radial velocity, $v_r$ goes to $0$ in the first orbit). Eventually these dissociated stars settle into smaller orbits with a constant envelope. 



The middle and last rows of Figure~\ref{fig:stellar_history} show the spatial and velocity history of stellar particles that exist in the intermediate radial range of [100, 105] ckpc/$h$ at $z=0$ and a larger radius [300, 310] ckpc/$h$ at $z=0$. Even at these radii we see two populations of particles -- those that have been accreted before $z=1$ and are in stable orbits, and those that have been accreted with a subhalo. Additionally, we also note that the largest radii (bottom row) have particles with high angular momentum and have never been in the center of the halo, with orbits always above 40 ckpc/$h$. The dissociation trends are similar to the previous cases.




\subsubsection{Dissociation and Splashback Distances}

\begin{figure}
    \includegraphics[width=1.0\columnwidth]{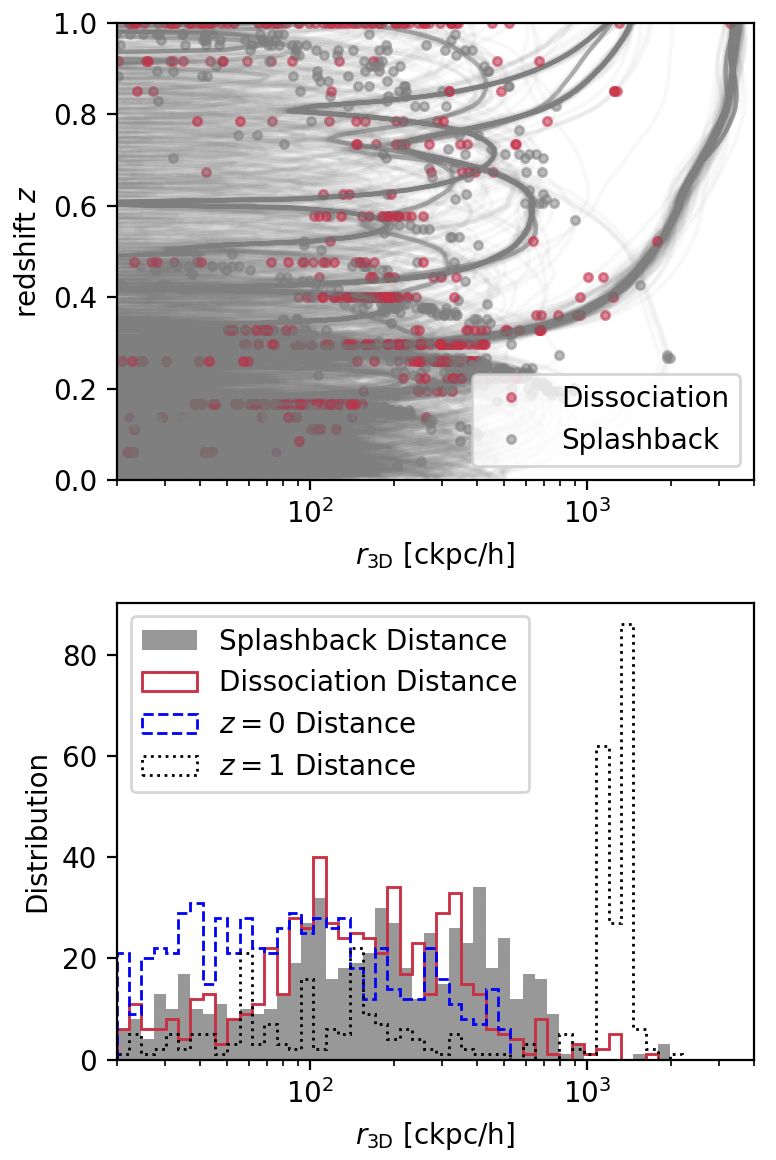}
    \caption{Evolution history of a random sample of stellar particles with their $z=0$ distances between 20 and 500 ckpc$/h$ in the TNG50-1 simulation. Upper panel: the particles' radial distances to the cluster center {\it vs} redshifts, noting where they become dissociated from subhalos, as well as their splashback distances (the furthest distances they reach during their first orbit) after dissociation. Lower panel: distributions of the particles' dissociation distances, splashback distances, as well as their $z=0$ and $z=1.0$ distances. The particle's dissociation and splashback distances first peak around 100 ckpc/$h$.}
    \label{fig:stellar_history_v2}
\end{figure}

In the previous subsection, we note a trend that many diffuse stellar particles of the cluster become dissociated from the subhalos only recently and in intermediate-radius orbits. They settle into stable orbits shortly after the dissociation. We further investigate if there exists a preferential radius where these particles become dissociated or "splashback" (reaching the first apocenter after dissociation) around the cluster center.

For this investigation, we randomly sample $\sim$1000 stellar particles in the radial range of [20, 500] ckpc/$h$ at $z=0$, and track their evolutionary history in different redshift snapshots. We also determine a "splashback" distance for these stellar particles -- defined as the furthest radius they reach before their radial velocities change signs, but after their dissociation from subhalos or $z=1$ whichever is later. 
We note that the particle's splashback and dissociation distances tend to be close, the distribution of these distances is plotted in Figure.~\ref{fig:stellar_history_v2}. A potential explanation is that the particles start dissociating after first pericenter passage and separate from the subhalo by splashback. Since the maximum distance between the stellar particle and the subhalo is at splashback, the dissociation time often coincides. 

Figure ~\ref{fig:stellar_history_v2} shows the evolution in distance for the particles. It also shows the distributions of their dissociation distances, their splashback distances, as well as their $z=0$ and $z=1$ distances. Their $z=1$ distance distribution shows a prominent peak outside of 1000 ckpc/$h$, indicating that many of these particles have been accreted from far away.  At $z=0$, the distances are skewed towards the cluster center, reflecting the radial density profile of the CG and ICL -- most dense towards the cluster center, and gradually thinning out at large distances.

Notably, the distributions of the particles' dissociation and splashback distances are {\bf not} skewed towards the cluster center. Their distributions offset significantly from the particles' $z=0$ distances. The shape of the distributions start to deviate from a flat distribution around 70 ckpc/$h$ and reach the first peak around 100 ckpc$/h$. We notice several further peaks in the dissociation and splashback distributions, most notably at 200 and 300 ckpc$/h$, which may be associated with different accretion events. These kinds of distributions can only happen if the dissociation and splashback process are happening much more frequently at larger distances, starting around 70 ckpc$/h$ and extending to several hundreds of kiloparsecs. 

We conclude here that the dissociation and splashback of stellar particles are more likely to happen beyond 70 ckpc$/h$. The density minimum we detected in data may indicate the extent of the active accretion region, and the transition between the CG and cluster diffuse light, where stellar particles are stripped from subhalos, and complete their first orbit around the cluster center.

\subsection{Stacked Profile}
\label{sec:stacked_sim}

\begin{figure}
    \includegraphics[width=1.0\columnwidth]{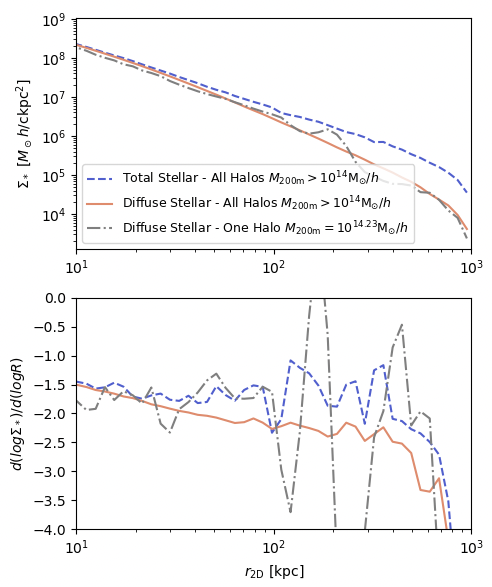}
    \caption{Upper panel: stacked stellar density profiles of galaxy cluster-sized halos in the TNG300-1 simulation, projected onto the $X$-$Y$ plane. We show the total stellar density profiles, and the diffuse component. The diffuse stellar density profiles are smoother than the total stellar profile. Lower panel: Radial slope profiles of the diffuse stellar component and the total stellar component of the clusters. While the slope of the diffuse stellar component becomes steeper with increasing radius, we do not observe a notable dip feature, unlike our measurements with the data. We also examine the profiles of one single halo in TNG300-1 with mass ($M_\mathrm{200} \sim 10^{14.23} M_\odot/h$) comparable to the TNG50-1 simulation, shown as the dot-dashed lines in both panels, which similarly show a relatively flat diffuse profile no consistent prominent dip (other than measurement noises).}
    \label{fig:sims_profile}
\end{figure}

The previous subsections are focused on the accretion features of one massive galaxy cluster in the TNG50-1 simulation. In this subsection, we use the TNG300-1 simulation which has a lower baryon resolution, but has 254 galaxy clusters of $M_\mathrm{200} > 10^{14} M_\odot/h$ at $z=0$ (with an average mass of $2\times 10^{14} M_\odot/h$, i.e. slightly higher than the mass of the TNG50-1 cluster) \footnote{We exclude halos within 1 cMpc/$h$ of the simulation boundaries.} to derive higher signal-to-noise measurements. Figure~\ref{fig:sims_profile} shows the stacked stellar density profiles and the slopes of these massive halos. 

We separate the stellar components into the CG+ICL and subhalo components as previously described. The stellar mass of the CG and ICL dominates the cluster's total stellar mass within $\sim$ 30 kpc/h, after which the subhalo stellar mass starts to dominate the cluster's total stellar mass. 
From the radial gradient profiles, the diffuse stellar profile becomes slightly steeper (more negative) with increasing radius within 100 ckpc/h, like what has been observed in the data, and in the TNG50-1 simulation. 

However, the similarity between the simulation and observations are rather limited. For example, the steepening slope with the TNG300-1 simulation is more mild than in data. Additionally, there is no prominent dip feature in the stacked profile. Although, the single halo profile does show the accretion features corresponding to density minima, it appears these features are washed out in the stacked profiles of clusters in TNG300-1. We also examine the profiles of one single halo in TNG300-1 with mass ($M_\mathrm{200} \sim 10^{14.23} M_\odot/h$) comparable to the TNG50-1 simulation, which similarly show a relatively flat diffuse profile with no prominent dips (the dot-dashed lines in Figure~\ref{fig:sims_profile}). We speculate that the lack of the dip-like features in TNG300-1 simulation maybe due to its significantly lower resolution compared to TNG50-1 (see Section~\ref{sec:appen_res} of the Appendix for more analysis). It is possible that baryon simulations with much higher resolutions are needed to fully reveal the accretion features of diffuse light in galaxy clusters. 

We also note here that in the Three Hundred hydrodynamical simulation, when investigating the 3D diffuse stellar distributions of the most massive halo in the simulation, \cite{2025arXiv250807232W} has revealed a similar steepening feature in the outer region of the CG.

\section{Summary and Discussion} \label{sec:discussion}

In this paper, we discuss the detection of a splashback-like steepening in the surface brightness profile of CG and ICL of DES redMaPPer clusters. The dip occurs between 40 to 60 kpc, with a mild dependence on cluster richness, and potentially indicates the beginning of the splashback-zone at the CG outskirts, where stellar particles are at outermost apocenter. We study simulations to gain theoretical insights into the phenomenon. The diffuse stellar content of a galaxy-cluster sized halo in the TNG50-1 simulation contains accretion features such as shell-like structures in their phase space and density enhancements at the apocenters of these shells, that appear as features in the radial logarithmic slope profiles of density. 
The simulation studies also revealed that diffuse stellar particles in the CG and ICL at $z=0$ can be accreted far away from the halo center. These particles often orbit around the cluster at an intermediate radius first, and in the process dissociate from subhalos, splashback towards the cluster center, and then settle into a closer orbit. These effects appear to occur preferentially at 70 ckpc$/h$ and beyond the center of the cluster.

Note that many literature studies, using simulations and observations, have found shells, tidal tails, or stellar stream features to be prevalent around massive galaxies \citep[e.g.,][]{2018MNRAS.480.1715P, 2018ApJ...866..103K, 2024MNRAS.530.4422K, 2024A&A...686A.182V, 2025MNRAS.543.3391K, 2025MNRAS.541.3015S, 2024ApJ...974..299Y}, which helps us understand past accretion and merging events \citep[e.g,][]{2019A&A...632A.122M, 2019MNRAS.487..318K, 2020MNRAS.498.2766W, 2022A&A...660A..28B,2024MNRAS.529..810R, 2024OJAp....7E..93E, 2025A&A...700A.104U, 2026MNRAS.545f1989K}. \cite{2025MNRAS.541.3015S} found that these features occur around 36\% of the nearby massive galaxies studied with CFHT observations, while the simulation and modeling study by \cite{2025MNRAS.543.3391K} found that these features can survive for $3\pm2$ Gyrs within galaxy clusters. These studies corroborate our speculation of an active accretion region at the CG outskirt causing splashback-like phase-space and density enhancements.

The splashback-like steepening feature measured in this paper agrees with the detection in a single galaxy cluster in \cite{2021MNRAS.507..963G}, described in the introduction. However, we note that the galaxy cluster in \cite{2021MNRAS.507..963G}, MACS J1149.5+2223,  has a mass of  $3.4\pm0.7\times10^{15} M_\odot$ \citep{2014ApJ...795..163U} and is much more massive than the average DES redMaPPer clusters, which would explain its larger dip radius measured at 70 kpc. The redMaPPer clusters in the richness bin $[60, \infty]$ in Figure~\ref{fig:richness_redshift} have an average mass of $10^{14.9} M_\mathrm{\odot}/h$ \citep{2019MNRAS.482.1352M}, which is  $\sim1/4.3$ of the mass of MACS J1149.5+2223. According to Equation~\ref{eq:richness} and interpreting richness dependence as mass dependence, the dip radius of MACS J1149.5+2223 would be $\sim$1.37 times the measurement in the highest richness bin in this paper.

In addition to the slope measurements discussed in \cite{2021MNRAS.507..963G}, many other observational studies have noted slope transitions in the surface brightness measurements of CG and ICL  \citep[see, for example, studies in][]{2016ApJ...820...42I, 2021MNRAS.508.2634Y, 2021ApJ...910...45M, 2024A&A...689A.306S, 2025A&A...697A..13K}, despite not explicitly computing the slope profiles. In fact, \cite{2021MNRAS.507..963G} already discussed such a feature in the precursor measurements used in this paper \citep{2019ApJ...874..165Z}. 
Although this density slope transition feature is commonly observed, we note that it may not always be prominent in every galaxy cluster \citep[for example, some of the profiles presented in][]{2020ApJS..247...43K}.

In addition to slope measurements, the transition of many other properties has been observed in this dip region, including diffuse light color and globular cluster densities, as demonstrated in \cite{2025A&A...697A..13K}. Most notably, kinematic studies have detected a clear distinction between central galaxies and surrounding low-surface brightness features, in particular, the stark difference between the velocity dispersion slopes of normal elliptical galaxies and the brightest cluster or group galaxies (BCGs). The majority of elliptical galaxies have falling velocity dispersion profiles \citep{2001AJ....121.1936G, 2004NewA....9..329P, 2006MNRAS.366.1126C}. However, BCGs are different. If observed out to large enough radii, most show flat or rising profiles \citep{2013ApJ...765...24N,2018MNRAS.477..335L,2024MNRAS.530.3924E}. In some cases, the velocity dispersion has even been measured to reach the value of the velocity dispersion of the cluster itself. This has been seen in both diffuse stellar spectra \citep{2015ApJ...807...56B,2020MNRAS.491.2617E} and in using globular clusters as kinematic tracer particles \citep{2011A&A...531A.119R}. \citet{2018MNRAS.477..335L} found that the velocity dispersion slope was positive only if the BCG was bright in the K-band, had a larger central velocity dispersion, and if the cluster also had a velocity dispersion greater than 600$\,$km/s. Simulations can explain this behavior as more massive galaxies have more stars that formed ex-situ, and thus larger stellar dispersions \citep{2023MNRAS.520.5651C, 2024ApJ...968...96H}.  \citet{2020MNRAS.491.2617E}, found that BCG slopes were positive when the velocity dispersion of the cluster dominated over that of the galaxy ($\sigma_0/\sigma_{clus}<0.5$) and were negative if the reverse was true ($\sigma_0/\sigma_{clus}>0.5$). Assuming the measurement traces the underlying gravitational field, this means the slopes are positive when the cluster signal dominates over that of the CG.

In terms of tracing distinct kinematic features in diffuse light around central galaxies, this has been done locally in Virgo.  In M87, \citet{2015A&A...579L...3L}  use Planetary Nebula (PN) to map out distinct kinematic structures in the cD galaxy's halo which are associated with a surface brightness substructure, presumed to be from an accretion event. \citet{2018A&A...616A.123H} likewise use PN to determine dynamically distinct structures in M49 and the intragroup light (IGL) of Virgo. They noted 50$\,$kpc as a transition to a higher fraction of IGL PN. \citet{2021ApJ...915...83T} have studied kinematics of globular clusters (GC) within 100$\,$kpc of M49's core. These authors noticed kinematic features in projected phase-space associated with shells. Azimuthally averaged density profiles of red intracluster GCs around M87 show a dip feature near 100$\,$kpc from the galaxy's core \citep{2018ApJ...864...36L}.

Studies like those above, which show distinct kinematic structures in the diffuse light around CGs, point to mergers and galaxy interactions as being important components to building up the diffuse light around CGs. Further, identifying velocity dispersion profiles that rise all the way to the cluster velocity dispersion value, suggest that the diffuse starlight follows the gravitational potential of the cluster, rather than the CG. Taken together, this is consistent with the notion that mergers and galaxy interactions can result in a diffuse stellar component that is distinct from the CG. 

Moving forward, it would be interesting to study the connection between the splashback-like dip and the kinematic features of the diffuse light around the CG. This splashback-like dip may also be related to the cluster's relaxation status, which is worth further exploration in a future paper.
The dip measurements presented here may provide an additional method for separating the stellar components \citep{2010MNRAS.405.1544D, 2011ApJ...732...48R, 2014A&A...565A.126P, 2021ApJS..252...27K, 2022ApJ...928...99C, 2022MNRAS.514.3082M, 2024A&A...692L...9C} of the CG and the diffuse light, which is often difficult to achieve in photometric studies. 
In addition, the diffuse light distribution is known to be correlated with the dark matter distribution in galaxy clusters \citep[e.g.,][]{2018MNRAS.474..917M, 2019MNRAS.482.2838M, 2021MNRAS.501.1300S, 2023A&A...679A.159D, 2024ApJ...965..145Y, 2024A&A...683A..59C, 2025ApJ...988..229Y, 2025MNRAS.539.2279B}, as well as total cluster mass \citep[e.g.,][]{2018MNRAS.480..521H, 2020MNRAS.491.3751D, 2025MNRAS.538..622G, 2023MNRAS.521..478G}. Connecting this splashback-like dip feature with the evolution history of the dark matter halo and its subhalos may help to further the understanding of the formation of cluster diffuse light, and thus better understand the galaxy cluster evolution process overall with upcoming cosmic surveys like the Rubin C. Observatory's Legacy Survey of Space and Time (LSST), and the Nancy Grace Roman Space Telescope.

\section*{Acknowledgments}

The work of YZ is supported by NOIRLab, which is managed by the Association of Universities for Research in Astronomy (AURA) under a cooperative agreement with the U.S. National Science Foundation. SA was supported by ANRF grant SRG/2023/001563 under the Government of India.
LOVE also acknowledges support from an NSF RUI award, 2205976 and LSST-DA Catalyst Fellowship 62192 through the support the John Templeton Foundation to LSST-DA. This material is based in part upon work while LOVE served at the National Science Foundation. Any opinions, findings, and conclusions or recommendations are solely those of the authors, and do not necessarily reflect the views of the National Science Foundation or the Federal government, LSST-DA or the John Tempelton Foundation.  We thank the anonymous referee for the very helpful suggestions and ideas that improved the quality of this work.

Some of the data underlying this article were accessed from the Illustris-TNG database. The IllustrisTNG simulations were undertaken with compute time awarded by the Gauss Centre for Supercomputing (GCS) under GCS Large-Scale Projects GCS-ILLU and GCS-DWAR on the GCS share of the supercomputer Hazel Hen at the High-Performance Computing Center Stuttgart (HLRS), as well as on the machines of the Max Planck Computing and Data Facility (MPCDF) in Garching, Germany. 

This project used public data products from the Dark Energy Survey (DES). Funding for the DES Projects has been provided by the U.S. Department of Energy, the U.S. National Science Foundation, the Ministry of Science and Education of Spain, the Science and Technology FacilitiesCouncil of the United Kingdom, the Higher Education Funding Council for England, the National Center for Supercomputing Applications at the University of Illinois at Urbana-Champaign, the Kavli Institute of Cosmological Physics at the University of Chicago, the Center for Cosmology and Astro-Particle Physics at the Ohio State University, the Mitchell Institute for Fundamental Physics and Astronomy at Texas A\&M University, Financiadora de Estudos e Projetos, Funda{\c c}{\~a}o Carlos Chagas Filho de Amparo {\`a} Pesquisa do Estado do Rio de Janeiro, Conselho Nacional de Desenvolvimento Cient{\'i}fico e Tecnol{\'o}gico and the Minist{\'e}rio da Ci{\^e}ncia, Tecnologia e Inova{\c c}{\~a}o, the Deutsche Forschungsgemeinschaft, and the Collaborating Institutions in the Dark Energy Survey.
 
The Collaborating Institutions are Argonne National Laboratory, the University of California at Santa Cruz, the University of Cambridge, Centro de Investigaciones Energ{\'e}ticas, Medioambientales y Tecnol{\'o}gicas-Madrid, the University of Chicago, University College London, the DES-Brazil Consortium, the University of Edinburgh, the Eidgen{\"o}ssische Technische Hochschule (ETH) Z{\"u}rich,  Fermi National Accelerator Laboratory, the University of Illinois at Urbana-Champaign, the Institut de Ci{\`e}ncies de l'Espai (IEEC/CSIC), the Institut de F{\'i}sica d'Altes Energies, Lawrence Berkeley National Laboratory, the Ludwig-Maximilians Universit{\"a}t M{\"u}nchen and the associated Excellence Cluster Universe, the University of Michigan, the National Optical Astronomy Observatory, the University of Nottingham, The Ohio State University, the OzDES Membership Consortium, the University of Pennsylvania, the University of Portsmouth, SLAC National Accelerator Laboratory, Stanford University, the University of Sussex, and Texas A\&M University.
Based in part on observations at Cerro Tololo Inter-American Observatory, National Optical Astronomy Observatory, which is operated by the Association of Universities for Research in Astronomy (AURA) under a cooperative agreement with the National Science Foundation.

\section*{Data Availability}

The data products and code used in this analyses can be found at \url{https://github.com/yyzhang/BCG_splashback_2026.git} and on Zenodo: doi: 10.5281/zenodo.20044368.

\appendix
\section{Simulation Resolution}
\label{sec:appen_res}

\begin{figure*}
    \includegraphics[width=1.0\columnwidth]{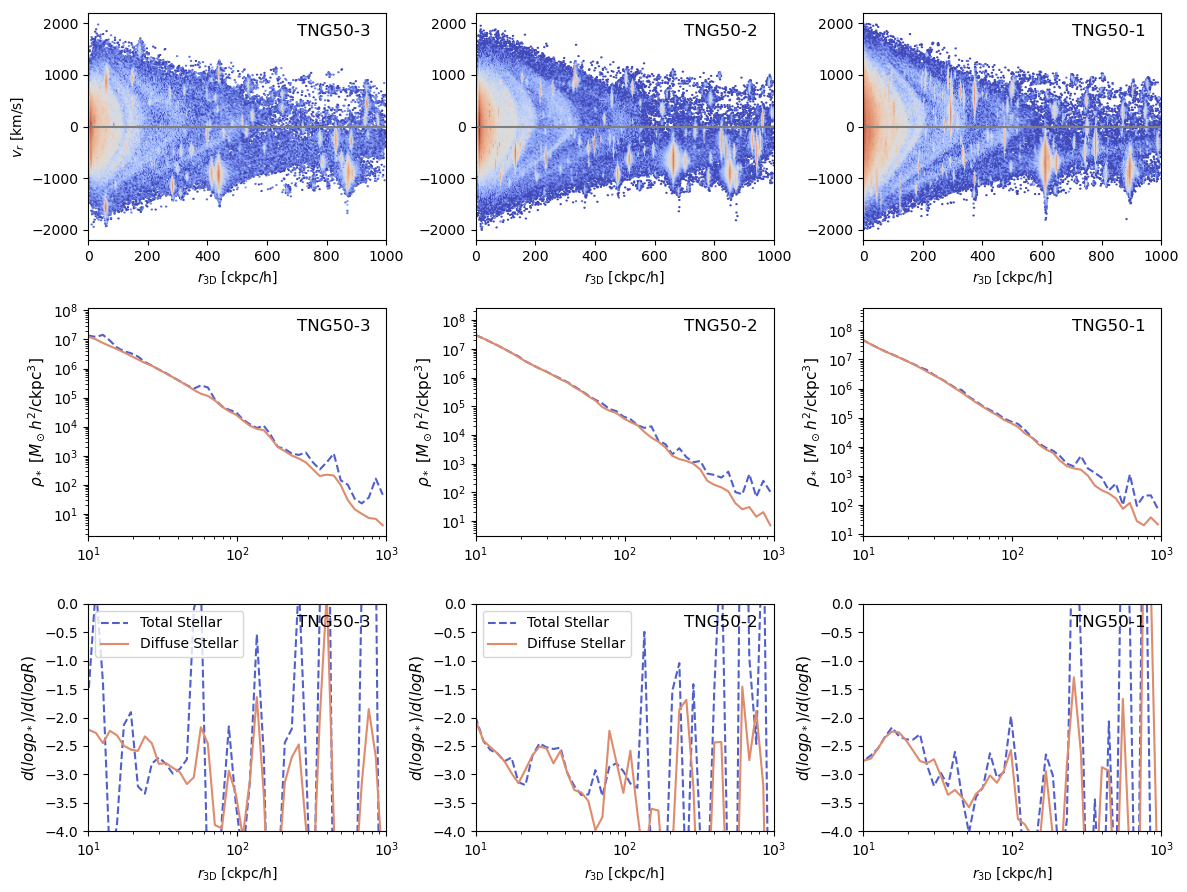}
    \caption{Upper row: Phase-space diagram (radial velocity {\it vs} radial distance) of the cluster (diffuse and subhalo) stellar particles in TNG50-3, TNG50-2 and TNG50-1 with increasing resolutions from left to right. The shell-like structures are most prominent and ``sharp'' (with clear edges) in the highest resolution simulation TNG50-1. Middle row: 3D  density profiles of the cluster stellar contents. Lower row: the radial slopes of the 3D density profiles. Density transition features are most evident in in the highest resolution simulation TNG50-1, but more difficult in the TNG50-3 simulation.}
    \label{fig:sim_resolution}
\end{figure*}

In Section~\ref{sec:stacked_sim}, we note that the splashback-like dip feature is not seen in stacked cluster density profiles from the TNG300-1 simulation. We speculate that the lack of this feature in TNG300-1 may be due to the significantly lower resolution compared to the TNG50-1 simulation. 

We further investigate the effects of simulation resolution with the TNG50-3, TNG50-2 and TNG50-1 simulations, in the order of increasing resolutions, as shown in Figure~\ref{fig:sim_resolution}. The dark matter mass resolution, $m_\mathrm{DM}$, for each of the three simulations is $2.9\times 10^7$, $3.6\times 10^6$ and $4.5\times 10^5$ $M_\odot$ respectively, while the gas mass resolution $m_\mathrm{gas}$ for each is $5.4\times10^6$, $6.8\times10^5$ and $8.5\times10^4$ $M_\odot$. Note that the resolution of TNG50-3 is still higher than the TNG300-1 simulation ($m_\mathrm{DM}=5.9\times 10^7M_\odot$, $m_\mathrm{gas}=1.1\times 10^7M_\odot$).  Figure~\ref{fig:sim_resolution} shows the phase space diagram of the stellar particles in the one-cluster sized halo in these TNG50 simulations, their radial density profiles and the gradient of these profiles. While the density transition features in the density gradient profiles and the shell features in the phase-space diagram are evident and ``sharp''(with clear edges) in TNG50-1, they are less evident in TNG50-2. In TNG50-3, the density gradient transition features are much more difficult to discern. We speculate that to identify these transition features in the simulations, one may need a simulation resolution at the level of TNG50-2 or better.


\bibliography{example}{}
\bibliographystyle{aasjournalv7}

\end{document}